\documentclass[12pt,thmsa]{article}
\usepackage{sw20aip}

\input{tcilatex}

\begin{document}

\title{Understanding quantum effects from classical principles}
\author{Nenad Klipa \\
Department of Physics, University of Zagreb, Bijeni\v cka\\
32, Zagreb, Republic of Croatia \and S. Danko Bosanac \\
Institut R. Bo\v skovi\' c, Zagreb, Republic of Croatia}
\date{\today}
\maketitle

\begin{abstract}
Dynamics of a particle is formulated from classical principles that are
amended by the uncertainty principle. Two best known quantum effects:
interference and tunneling are discussed from these principles. It is shown
that identical to quantum results are obtained by solving only classical
equations of motion. Within the context of interference Aharonov-Bohm effect
is solved as a local action of magnetic force on the particle. On the
example of tunneling it is demonstrated how uncertainty principle amends
traditional classical mechanics: it allows the momentum of the particle to
change without the force being the cause of it.
\end{abstract}

\section{Introduction}

The phrase \textit{quantum effect} was introduced into the language of
physics to signify everything that cannot be understood in terms of
traditional classical mechanics. It is used to emphasize deep division
between the two views: one of the classical and the other of the quantum
world. From the failures to explain them from classical principles a very
important conclusion was reached: quantum dynamics is fundamental and
classical is derived in the limit $\hbar \rightarrow 0$. This view can be
relaxed somewhat by accepting that the approximation $\hbar \sim 0$ is
sufficient for classical dynamics to be a suitable alternative to quantum,
and it is called semi-classical. In this approximation one combines
classical and quantum principles, but certain caution should be exercised
because deciding when it is applicable is not sometime clear. Quantum
effects in this approximation are partially described, and this is the
closest to what one can achieve with the classical principles.

Which are the quantum effects? In the first place this is interference and
tunnelling, but there are many more: zero point energy, spin, discrete bound
states, photo-electricity, ... to mention only few. As it was mentioned
classical mechanics could not explain their physical origin from its
principles, but the question is does the quantum mechanics? It does, but in
saying so one thing should not be overlooked: in order to explain them the
wave-particle dualism must be accepted without being able to rationally
derive it from other principles. In fact the problem is not to comprehend
either of the components: the concept of particle is well understood and so
is the concept of the wave, but the union of the two. However, by accepting
this logical system, the quantum mechanics, in which the wave-particle
dualism is the postulate (and few others), one derives equation the
solutions of which describe all the phenomena (not entirely, e.g. to
describe the spontaneous emission one needs more elaborate theory). While
there is no problem with the mathematical aspect of obtaining the correct
result, there are problems with the physical understanding of them. However,
the epithet "physical" must be defined. Within the logical system of quantum
mechanics "physical" means that the essential features of the phenomena can
be described by the properties of either the waves or the particles. In this
respect the physics of interference is easily described by the wave-like
nature of particles, but, for example, when it comes to the effect of
tunneling there are problems. It cannot be explained from the particle-like
nature of particles, but neither it can be from the wave-like nature. In
other words, there is no explanation how the waves get through the potential
barrier, except that the solution of the wave equation predicts it.

Are there any alternative formulations of quantum mechanics that is based on
the wave-particle dualism? This question inevitable rises the following one:
why there is necessity for alternative formulations? The answer to the
second is that alternative formulations offer a different viewpoint of
physics, and as such they are very important, and the answer to the first is
that there are. One alternative formulation is based on strictly abstract
approach, in which the essential postulate is that there is
observable-operator connection \cite{neumann}. Apart from that one there are
other alternative formulations of quantum mechanics. The path integral
method is the best known \cite{feynman,hibbs}, in which instead of
Schroedinger equation one postulates its integral equation form. The other
is the Bohm`s method of quantum potentials \cite{bohm}, which in essence is
not new formulation because classical trajectory equations are solved in the
effective potential that is obtained from solution of Schroedinger equation.
The third formulation is random classical mechanics \cite{nelson}, which
departs from the usual approach to quantum mechanics by introducing the
concept of probability into classical mechanics, however Schroedinger
equation is retained indirectly by postulating imaginary diffusion
coefficient for the probability. Characteristic of all the alternative
formulations is that Schroedinger equation is postulated, in one form or the
other, while classical mechanics plays no important role.

Alternative formulations are also known in classical mechanics, for example
Lagrange or Hamilton formulations \cite{goldstein}. They do not offer, in
essence, anything different than the Newton equations of motion, but in many
circumstances are more useful and emphasize different concepts in physics
(e.g. the energy conservation law). However, one obvious formulation of
classical mechanics had been entirely neglected. For centuries the basic
ingredient of the scientific method was the concept of error, because no
result of experiment is considered reliable if the error margin is not
cited. Yet to the best of knowledge no reference work undertook to discuss
the following question: given the error margins for initial conditions how
this error propagates in time? This question is of utmost importance for
theoretical predictions, because it can be shown (but not elaborated in
details here) that the assumption of the precise initial conditions is
academic in most circumstances. The meaning of this is that even the tiniest
error, often in a relatively short time, increases to such an extent that no
reliable predictions could be made. Therefore instead of asking whereabouts
of the particle if its initial conditions are known, more appropriate
question is to ask for probability of whereabouts of the particle if the
probability of its initial conditions is known. This shift in emphases means
that instead of treating dynamics of a point in the phase space one should
treat dynamics of a phase space density. In other words, the emphases is
shifted from the Newton equations of motion to the Liouville equation.
Again, the two formulations are equivalent, although analysis of the phase
space density provides additional insight into the dynamics of the system.

Importance of formulating classical dynamics in terms of the Liouville
equation is that the uncertainty principle can be imposed on its solutions,
and this condition can be treated as the additional postulate in classical
mechanics \cite{bos3}. It says that if the standard deviations for the
coordinate and the momentum are $\Delta x$ and $\Delta p$, respectively,
then 
\begin{equation}
\Delta x\;\Delta p\succeq c  \label{un}
\end{equation}
where $c$ is a constant (determination of the constant, which has the value $%
c=\hbar /2,$ is not discussed, but it can be done in the same way as from
the black-body radiation law, by fitting theoretical predictions to the
experimental data). By implementing the postulate it remains to find
solution of the Liouville equation with that property. It is anticipated
that quantum mechanical results will be obtained, and if this is the case
then this approach could be treated as alternative formulation of quantum
mechanics. The most important difference with the previous formulations is
that the starting point is classical dynamics, which is formulated with the
Liouville equation but amended with the uncertainty principle. The other
formulations start, in one form or the other, by postulating Schroedinger
equation, however, if the suggested formulation is correct Schroedinger
equation should be derived. Therefore, this approach does not replace
quantum mechanics but derives it from different principles that do not
incorporate the wave-particle dualism. It should be pointed out that the
uncertainty principle in quantum mechanics is the law, i.e. it is derived
from more basic principles, e.g. the wave-particle dualism, while in this
formulation it acquires the status of postulate.

The problem of implementing the uncertainty postulate into classical
mechanics is purely a mathematical task. The main problem is to find a
suitable parametrization of the phase space density that ensures that the
amendment is fulfilled at all times. The solution was demonstrated in
several instances \cite{bos4,bos5,bos6,bos7}, but for the sake of
completeness it will be described in the following section. Once this is
done then the time evolution of the probability densities is obtained by
solving the Liouville equation, but this essentially means solving Newton`s
equations of motion. This is the essence of what it will be called the 
\textit{classical solution} for dynamics of particles. In short, suitable
parametrization of the phase space density that ensures the uncertainty
principle, plus the Newton's equations of motion, is \textit{classical
dynamics}. In contrast the \textit{traditional classical dynamics} is based
on the concept of trajectory and without the uncertainty principle. \textit{%
Quantum dynamics} implies starting from the same initial conditions as in 
\textit{classical dynamics}, but solving Schroedinger equation instead of
Newton's. As it will be shown classical dynamics describes two very
important quantum effects: interference and tunneling (in this context the
problem of the zero point energy will also be discussed).

\section{Formulation of classical dynamics}

\label{sec:theory}The suggestion in Introduction of incorporating the
uncertainty principle into classical mechanics seems contradiction with the
concept of trajectory, the concept that is an integral part of traditional
classical mechanics. This is indeed correct but it is no longer that if
instead of deterministic view one assumes the probabilistic one. The
arguments for this change were mentioned in Introduction. The shift of
emphases in classical mechanics from the concept of trajectory to the
concept of probability means that formally one replaces the Newton equations
of motion 
\begin{equation}
\frac{d\vec{p}}{dt}=\vec{F}\quad ;\quad \frac{d\vec{r}}{dt}=\frac{\vec{p}}{m}
\label{New}
\end{equation}
with the Liouville equation for the probability density in the phase space 
\begin{equation}
\frac{\partial \rho }{\partial t}\,+\,\frac{\vec{p}}{m}\cdot \nabla _{r}\rho
\,+\,\vec{F}\cdot \nabla _{p}\rho \;=\;0  \label{liouv}
\end{equation}
where $m$ is mass of particle and $\vec{F}$ is force acting on it. One step
towards implementing the uncertainty principle (\ref{un}) is to change the
meaning of the function $\rho $. Instead of being treated as the probability
density one should accept that it is a general, but real, function that
satisfies the Liouville equation. The reason for this change is the
observation that $\rho $ cannot be measured accurately because that implies
accurate measurement of both the position and momentum of a particle, and
this would violate the uncertainty principle. On the other hand, for the
averages 
\begin{equation}
P(\vec{r},t)\;=\;\int d^{3}p\,\,\rho (\vec{r},\vec{p},t)\;\;\;\;;\;\;\;\;Q(%
\vec{p},t)\;=\;\int d^{3}r\,\;\rho (\vec{r},\vec{p},t)  \label{pq}
\end{equation}
this restriction is not applicable because, for example, for the probability 
$P(\vec{r},t)$ to be measured it is not necessary to know the momentum.
Therefore the phase space density (not the probability density) is treated
as an auxiliary function that satisfies the Liouville equation, and whose
initial value is obtained from the quantities such as (\ref{pq}), or from
the probability current 
\begin{equation}
\vec{J}(\vec{r},t)\;=\;\frac{1}{m}\int d^{3}p\,\,\vec{p}\;\rho (\vec{r},\vec{%
p},t)\;  \label{curr}
\end{equation}
Therefore the quantities that have physical significance are the
probabilities (\ref{pq}) and the probability current (\ref{curr}), and not
the phase space density $\rho $, although the time evolution of the former
is derived from the latter.

The uncertainty principle requires that $P(\vec{r},t)$ and $Q(\vec{p},t)$
are related by the inequality (\ref{un}), which puts a constraint on the
possible phase space densities, the solutions of Liouville equation. The
problem is, therefore, to select the family of functions with that
requirement, which can be readily solved if certain rules from the Fourier
analysis (for a reference see \cite{korner}) are recalled. According to
these rules the probability densities are written as 
\begin{equation}
P(\vec{r},t)\;=\;|f(\vec{r},t)|^{2}\quad ;\quad Q(\vec{p},t)\;=\;|g(\vec{p}%
,t)|^{2}  \label{pqq}
\end{equation}
and if the two functions are interrelated by 
\begin{equation}
f(\vec{r},t)\;=\;\frac{1}{\sqrt{\left( 2\pi \hbar \right) ^{3}}}\int dp\,e^{i%
\vec{p}\cdot \vec{r}/\hbar }\,g(\vec{p},t)  \label{ft}
\end{equation}
then the inequality (\ref{un}) is ensured. The relationships (\ref{pqq}) and
(\ref{ft}) are known from quantum mechanics, in fact they are integral part
of it. It would appear therefore that in this way quantum mechanics is
introduced through the ''back door'', however this is not correct. The
mentioned relationships are a mathematical way of selecting those
probabilities $P(\vec{r},t)$ and $Q(\vec{p},t)$ that obey the uncertainty
principle, and before being used in quantum mechanics they were known in the
Fourier analysis. The same relationships are also used in the signal theory,
although not for the probabilites but for the intensities of pulses in the
time and frequency domains.

The next step is to find how the phase space density is related to the
amplitude $f$, because in this way the required constraint on the solutions
of Liouville equation would be ensured. The relationship should be obtained
by using the definitions (\ref{pq}), which are familiar rules for
convolutions in Fourier analysis, and from that observation one obtains 
\begin{equation}
\rho (\vec{r},\vec{p},t)\;=\;\frac{1}{\left( \pi \hbar \right) ^{3}}\,\int
dq\,e^{2i\vec{p}\cdot \vec{q}/\hbar }\,f^{*}(\vec{r}+\vec{q},t)\,f(\vec{r}-%
\vec{q},t)  \label{w}
\end{equation}
In order to prevent possible misunderstandings few comments about the
function (\ref{w})\cite{wig1,wig2} are in order. It is known as the Wigner
function, but it should not be considered here as the Wigner
quasi-probability distribution i.e., the Weyl transform (up to constant) of
a pure state density operator, because we did not introduced any quantum
operator or state. Remember that $f$ is only an auxiliary function used for
the parametrization of the classical probability for the coordinate of the
particle. The Wigner function \cite{wig1} is one of the many
quasi-probability distribution functions invented to express quantum
mechanical averages in the classical (phase space) manner, which is not the
subject under discussion in the present work.

The phase space density should be solution of the Liouville equation (\ref
{liouv}), and if the constraint is the parametrization (\ref{w}) then one
derives the equation for $f$. It is straightforward to show \cite{dos1} that
for the polynomial potentials up to the second degree the equation that one
derives for $f$ is 
\begin{equation}
i\hbar \frac{\partial f}{\partial t}\;=\;-\frac{\hbar ^{2}}{2m}\Delta
f\,+\,V\,f  \label{schr}
\end{equation}
in which Schroedinger equation is recognized. Therefore for the potentials
of this kind the equation (\ref{schr}) is classical solution for the problem
of implementing the uncertainty principle. This means that the problems such
as free particle, charged particle in the homogeneous (in general a time
varying) electric or magnetic field, harmonic oscillator, etc. are exactly
described by the amended classical theory. However, it can be shown that the
same conclusion is valid in general \cite{bos8}, and the proof is based on
observation that any potential can be divided-up into the segments of
constant value. In each segment the classical solution that is based on
solving (\ref{schr}) is in order, and by conveniently adjusting the boundary
conditions between segments one derives again (\ref{schr}), where now the
potential $V$ is a general function. Consequently, free particle
trajectories are used for the time evolution of the phase space density, a
rather complicated procedure but in principle exact. However, if one is not
interested in the solution in the phase space, only in the coordinate
subspace, then it is sufficient to solve the equation (\ref{schr}), which is
a much simpler task. The price, which is paid for this benefit, is the loss
of information that the phase space provides, which will be demonstrated on
the following examples.

It would appear that the steps that were taken here are in the reverse order
as it is done when deriving the Liouville equation from Schroedinger
equation, but in what way they are different is discussed in the Conclusion
of this paper.

\section{Interference}

The best-known quantum effect is interference that results from a particle
having a choice to get through two slits. The setup is the following. In the
y-z plane there is a screen with two circular slits that are centered on the
y-axes at $\pm \;y_{0}$, having the width $\Delta $. The screen ends at $x=0$%
. The particle is sufficiently de-localized before the screen so that there
is equal probability to enter either of the slits. In the slits, and just
before exiting them, the probability of finding particle in the plane $x=0$
is a sum of the form $P(x,y,z)=P_{1}(y,z)g_{1}(x)+P_{2}(y,z)g_{2}(x)$, where
the index designates the probability centered at a particular hole. The
probability $P_{i}(y,z)$ is not zero in a circle of the radius $\Delta $,
and for simplicity it will be assumed to have the form of a Gaussian with
that width. The probability $g_{i}(x)$ is determined essentially by the
length of de-localization of the incoming particle along the x-axes, and for
simplicity it will also be assumed to be Gaussian of the width $\Delta $.
The average velocity of the particle in the x direction is $v_{0}$. The fact
that the particle enters each slit with the same probability implies that
the functional forms for $P_{1}(y,z)g_{1}(x)$ and $P_{2}(y,z)g_{2}(x)$ are
the same. The moment when the maximum of the probability $P(x,y,z)$ is at $%
x=0$ will be $t=0$. Propagation of this probability density from this moment
on will be as for a free particle, and the impact that the screen has on the
motion in $x>0$ will be neglected. Without considering the uncertainty
principle the motion of the probability density can only be deduced if in
addition to the probability $P(x,y,z)$ one also knows the velocity
distribution of the particle. In the traditional classical mechanics this
distribution is arbitrary, but with the uncertainty principle included it is
no longer that. In fact the velocity distribution is not the main problem,
it is the initial phase space density from which the initial conditions for
classical trajectories are selected. The phase space density, if the
uncertainty principle is included, is calculated from (\ref{w}) where the
function $f$ is defined by (\ref{pqq}). It is obvious that the definition (%
\ref{pqq}) does not determine this function uniquely, because $f$ is in
general complex. The phase of $f$ is determined from the probability current
(\ref{curr}), and if (\ref{w}) is replaced for the phase space density it
can be easily verified that 
\[
P\;\nabla Arg(f)=\vec{J} 
\]
By assuming that the probability current is known the phase is calculated as
the line integral of the function $\vec{J}/P$. In the example with the two
slits the current in the y-z plane is zero, while in the x direction it is
given by $J=v_{0}g(x)$.

In this way the initial conditions are determined and the function $f$ is 
\[
f(x,y,z)=\left[ \sqrt{P_{1}(y,z)}+\sqrt{P_{2}(y,z)}\right] \sqrt{g(x)}%
\;e^{iv_{0}x} 
\]
where from now on it will be assumed that $m=\hbar =1$ (the square root of a
sum of two functions is equal to the sum of the square roots of these
functions only if they do not overlap). For a particular example of the
Gaussian probabilities the function $f$ is (non-essential factors are
omitted for convenience) 
\[
f(x,y,z)=\left[ e^{-\frac{1}{2\Delta ^{2}}(y-y_{0})^{2}}+e^{-\frac{1}{%
2\Delta ^{2}}(y+y_{0})^{2}}\right] e^{-\frac{1}{2\Delta ^{2}}%
(x^{2}+y^{2})+iv_{0}x} 
\]
from which the initial phase space density is 
\begin{eqnarray}
\rho _{0}(x,y,z,v_{x},v_{y},v_{z}) &=&\left[ e^{-\frac{1}{\Delta ^{2}}%
(y-y_{0})^{2}}+e^{-\frac{1}{\Delta ^{2}}(y+y_{0})^{2}}+2\cos
(2v_{y}y_{0})e^{-\frac{1}{\Delta ^{2}}y^{2}}\right]  \label{ro2in} \\
&&e^{-\Delta ^{2}[(v_{x}-v_{0})^{2}+v_{y}^{2}+v_{z}^{2}]-\frac{1}{2\Delta
^{2}}(x^{2}+z^{2})}  \nonumber
\end{eqnarray}

At any time later, and if the particle is free, the phase space density is 
\begin{equation}
\rho (\vec{r},\vec{v},t)=\rho _{0}(\vec{r}-\vec{v}\;t,\vec{v})  \label{ro2t}
\end{equation}
from which the probability $P(\vec{r},t)$ is 
\begin{eqnarray*}
P(\vec{r},t) &=&\int d^{3}v\;\rho (\vec{r},\vec{v},t)= \\
&&\frac{1}{\left( \Delta ^{4}+t^{2}\right) ^{3/2}}\left[ \cosh \left( \frac{%
2\Delta ^{2}yy_{0}}{\Delta ^{4}+t^{2}}\right) +\cos \left( \frac{2tyy_{0}}{%
\Delta ^{4}+t^{2}}\right) \right] e^{-\frac{\Delta ^{2}\left[ \left(
x-tv_{0}\right) ^{2}+y^{2}+y_{0}^{2}+z^{2}\right] }{\Delta ^{4}+t^{2}}}
\end{eqnarray*}
The screen where the probability density is measured is at $x=X$ and as a
function of the z-y coordinates its typical form is shown in Figure 1 (left
pattern).

\FRAME{ftbpFU}{5.1145in}{3.1721in}{0pt}{\Qcb{Typical interference pattern
(left figure) from two slits that is observed on the screen along the
x-axes. The two slits are along the y-axes. The pattern is shifted (right
figure) if the localized magnetic field is placed in between the two slits.}%
}{\Qlb{fig1}}{fig1.wmf}{\special{language "Scientific Word";type
"GRAPHIC";display "USEDEF";valid_file "F";width 5.1145in;height
3.1721in;depth 0pt;original-width 7.0067in;original-height 4.3396in;cropleft
"0";croptop "1";cropright "1";cropbottom "0";filename
'fig1.wmf';file-properties "XNPEU";}}

The parameters where chosen arbitrarily, for the demonstration purpose only,
and their values are: $y_{0}=1000$, $\Delta =100$, and $v_{0}=1$. The screen
is located at the distance $X=10^{5}$, and the time at which the probability
is observed is $t=10^{5}$. Typical interference pattern is obtained, which
is the same as if the particle is treated as a wave. However, in the
treatment here the interference pattern is obtained by propagating the phase
space density by classical trajectories, the equation (\ref{ro2t}), and by
the classical rules of probability addition. This is made possible by having
additional term in the initial phase space density (the third term in (\ref
{ro2in})), besides those that correspond to the typical classical
probability densities that are centered around the slits. The interference
term in the phase space density, as the additional term can be called, has
two distinctive features: one is that it has both positive and negative
values, and the other that it is centered at a totally ''non-physical''
place, in between the two slits. The first feature is essential if by the
classical rules of addition of probabilities one can describe the
oscillatory structure of the probability density on the screen at $x=X$. The
negative values of the phase space density rule out the possibility to
attach to it physical significance of the probability density. However, as
it was mentioned, this ''non-physical'' character of the phase space density
is explained by impossibility to measure it in experiment.

While one could accept the possibility to work with the non-positive phase
space densities, the location of the interference term in between the two
silts rises at least two important questions. One is if it has physical
significance, because its location would imply that it is only a
''mathematical trick'' by which the correct result is obtained. The other
question is why is it placed in a region where it does not overlap with the
space where the particle is certainly located, around the slits?

The physicality of the interference term of the phase space density can be
tested, by applying the force on the particle that is only localized in the
region between the two slits. For example this can be homogenous magnetic
field that is localized in a tube of the radius smaller than $y_{0}$,
centered at $y=0$ and oriented in the z direction. In the traditional view
the phase space density has zero value in this region and hence the force
would not have any effect on the pattern on the screen at $x=X$. However,
the phase space density that is in accordance with the uncertainty principle
has a non-zero value around $y=0$ (the interference term), therefore all
trajectories that originate there are affected by the magnetic field (it is
tacitly assumed that the particle is charged). They will be affected for a
time $t=T$ until they exit the magnetic tube, and for simplicity it will be
assumed that this time is independent of the initial conditions of
trajectory. The equation for these trajectories is (for simplicity the mass
of particle is assumed to be unity) 
\[
d_{t}^{2}\vec{r}=\vec{v}\times \vec{H} 
\]
and for the assumed magnetic field, i.e. $\vec{H}=h_{0}\hat{z}$, the
solution is know exactly. After time $T$ the trajectory exits the magnetic
field, and after that it goes as if no force in applied on the particle. It
can be assumed that the magnetic field acts for a short time, meaning that $%
Th_{0}\ll 1$, in which case the trajectory is 
\[
\vec{r}=\vec{r}_{0}+\vec{v}_{0}t+\vec{v}_{0}\times \vec{H}\;\left( t-\frac{1%
}{2}T\right) 
\]

These trajectories are used to propagate the phase space density, but only
the interference part, because the trajectories that originate around the
two slits are not affected by the magnetic force. The time evolution of the
interference part is 
\[
\rho ^{(int)}(\vec{r},\vec{v},t)=\rho _{0}^{(int)}\left[ \vec{r}-\vec{v}t-%
\vec{v}\times \vec{H}\;\left( t+\frac{1}{2}T\right) ,\vec{v}+\vec{v}\times 
\vec{H}\;\right] 
\]
where $\rho _{0}^{(int)}$ is the third term in (\ref{ro2in}). The
probability $P(\vec{r},t)$ is now obtained by integrating the phase space
density over the momentum (velocity) variables, which is finally 
\begin{eqnarray*}
P(\vec{r},t) &=&\frac{1}{\left( \Delta ^{4}+t^{2}\right) ^{3/2}}\left[ \cosh
\left( \frac{2\Delta ^{2}yy_{0}}{\Delta ^{4}+t^{2}}\right) +\cos \left( 
\frac{2ty+h_{0}T^{2}x}{\Delta ^{4}+t^{2}}y_{0}\right) e^{-\frac{\Delta
^{2}T^{2}h_{0}v_{0}y}{\Delta ^{4}+t^{2}}}\right] \\
&&e^{-\frac{\Delta ^{2}\left[ \left( x-tv_{0}\right)
^{2}+y^{2}+y_{0}^{2}+z^{2}\right] }{\Delta ^{4}+t^{2}}}
\end{eqnarray*}
The effect of the localized magnetic field is real, despite the fact that it
only affects what appears to be ''non-physical'' part of the phase space
density. It is manifested essentially as the shift of the interference
pattern, as it is shown in Figure 1 (right pattern), where the parameters
are the same as before, but in addition $T=1000$ and $h_{0}=0.0002$.
Therefore, if the effect is confirmed then one can indeed argue that the
interference part of the phase space density is physically real.

The effect is known as the Aharonov-Bohm \cite{aharonov,silverman}, and its
main point was to show that the concept of potential is more physical than
the concept of force. This force-potential dilemma is historic, but it was
always thought that the latter is just a convenient mathematical
simplification of treating electromagnetic field. In quantum mechanics it is
the vector potential that enters the Schroedinger equation explicitly, and
not the magnetic force, and based on the Aharonov-Bohm effect it was argued
that with the concept of field one needs a non-local theory to explain it.
Namely, the magnetic field in this effect is confined in a localized region
in space while the vector potential is spread all over it, including both
slits. Therefore the latter has local effect on the charged particle, while
the magnetic force does not have. However, by starting from the classical
principles the force was reinstated as a legitimate concept, because it was
shown that the effect can be explained as a local event on the interference
term. It could be argued that the problem of non-locality with the concept
of field is replaced by the problem of non-locality in the phase space
density. After all, there is always a question (the second mentioned
earlier) why in the phase space there is a contribution that does not have
any direct relationship to where the particle indeed is, i.e. around one of
the slits? In fact the question is not so much why is it there (it is there
because of the uncertainty principle), but how it comes to be there?

In order to answer this question one needs to understand the meaning of the
initial conditions for the phase space density, and how they are formed. For
the two slit problem the initial probability was single centered, i.e. a
wide distribution that overlaps both slits. If the walls of the screen are
very thin, and for the particle they are infinitely high potential barrier,
then almost instantaneously the initial single centered distribution splits
into two, well separated ones. This means that before overlapping with the
screen the phase space density is single centered around $y=0$ and $p_{y}=0$
(the other degrees of freedom are not essential for discussion), but after
exiting the slits the phase space density is non-zero around three centers.
For the two, centered around $y=\pm y_{0}$, one can easily give arguments
why they are there. However, there exists the third, around $y=0$ (the
interference term), which is a surprise because it is behind the infinite
wall with respect to the original phase space density and it appears
instantaneously. The choice is now either to accept this fact, but then one
should accept the view that this classical approach is non-local theory, or
to argue that formation of the interference term takes time. Intuitively the
latter is the more acceptable view, but in order to prove it one would
really need to work with the relativistic theory, in which correlation
effects (and this is what one talks about in the formation of the
interference term) cannot travel faster than the speed of light. In other
words, sudden appearance of the interference term is nothing but an artifact
of non-relativistic theory, where the signals can travel at arbitrary speed.

Solving the relativistic two slit experiment is quite demanding, but without
the loss of generality one can treat simpler but analogous one dimensional
problem. At $t=0$ a probability distribution is formed around $y=0$ in such
a way that its momentum distribution contains two disjoined components, one
with the average momentum $-m_{0}$ and the other with $m_{0}$. It is
expected that the probability in the coordinate $y$ would split into two
components traveling in the opposite directions (provided that $m_{0}$ is
larger than the width of the momentum probability). After certain time one
would have two disjoined probability densities, of the sort as in the two
slits experiment, the only difference being that the two probabilities
travel in the opposite directions. By applying a force on each probability
distribution one can reduce their average momentum to zero, in which case
this would be precisely the initial conditions for two slits (one neglects
finer details that make this statement not entirely correct). This problem
can be treated by relativistic mechanics, in which case one observes how the
phase space density is formed.

The relativistic phase space density, from which Dirac equation is derived,
in single dimension is \cite{bos9} 
\begin{eqnarray*}
\rho (y,t,p,p_{0}) &=&\int dq\;dq_{0}\;\left[
f^{*}(y+q,t+q_{0})f(y-q,t-q_{0})+g^{*}(y+q,t+q_{0})g(y-q,t-q_{0})\right] \\
&&e^{2ipq-2ip_{0}q_{0}}
\end{eqnarray*}
where it was assumed that only positive energy components are present. At $%
t=0$ the phase space density is 
\begin{eqnarray*}
\rho _{0}(y,p,p_{0}) &=&\int dq\;dq_{0\;}\left[
f_{0}^{*}(y+q,q_{0})f_{0}(y-q,-q_{0})+g_{0}^{*}(y+q,q_{0})g_{0}(y-q,-q_{0})%
\right] \\
&&e^{2ipq-2ip_{0}q_{0}}
\end{eqnarray*}
where the functions $f_{0}$ and $g_{0}$ are defined in the momentum space as 
\[
f_{0}(y,q_{0})=\int
dk\;A(k)e^{iky-iq_{0}e(k)}\;\;\;;\;\;\;g_{0}(y,q_{0})=\int
dk\;w(k)A(k)e^{iky-iq_{0}e(k)} 
\]
$A(k)$ is the momentum space amplitude for the initial conditions (the units
are $m=c=\hbar =1$), $w(k)=k/[1+e(k)]$ and $e(k)=\sqrt{1+k^{2}}$ . By
evaluating the integrals in the variables $q$ and $q_{0}$ the initial phase
space density is (non-essential factors are omitted) 
\begin{eqnarray*}
\rho _{0}(y,p,p_{0}) &=&\int dk\;\delta \left[ 2p_{0}-e(p-k)-e(p+k)\right] %
\left[ 1+w(p-k)w(p+k)\right] \\
&&A^{*}(p-k)A(p+k)\;e^{2iky}
\end{eqnarray*}
and at any later time the phase space density is 
\begin{eqnarray*}
\rho (y,t,p,p_{0}) &=&\rho _{0}(y-\frac{p}{p_{0}}t,p,p_{0}) \\
&=&\int dk\;\delta \left[ 2p_{0}-e(p-k)-e(p+k)\right] \left[ 1+w(p-k)w(p+k)%
\right] \\
&&A^{*}(p-k)A(p+k)\;e^{2ik(y-\frac{p}{p_{0}}t)}
\end{eqnarray*}
The phase space density in only the coordinate-momentum variables is
obtained by integrating in the variable $p_{0}$ (the fourth component of the
four-momentum), in which case 
\begin{equation}
\rho (y,p,t)=\int dk\;\left[ 1+w(p-k)w(p+k)\right] A^{*}(p-k)A(p+k)%
\;e^{2iky-ie(p+k)t+ie(p-k)t}  \label{relro}
\end{equation}

For the momentum amplitude it is now assumed to have the form 
\[
A(p)=e^{-\frac{(p-m_{0})^{2}}{2\Delta ^{2}}}+e^{-\frac{(p+m_{0})^{2}}{%
2\Delta ^{2}}} 
\]
where $m_{0}>>\Delta $ (the two distribution do not overlap). When in the
phase space (\ref{relro}) the product of the amplitudes is evaluated one
gets three terms that are centered around $y=0$: two terms that are centered
around $p=\pm m_{0}$ and one around $p=0$. Between all three contributions
there is no overlap (or it is negligible). The first two contribute to the
phase space density that is centered around $y=\pm \frac{m_{0}}{e(m_{0})}t$
and $p=\pm m_{0}$, which is analogous to the phase space density centered
around $y=\pm y_{0}$ in the two slit setup. These two contributions recede,
each traveling at the speed $m_{0}$ (in the non-relativistic limit when $%
m_{0}\ll 1$) or at nearly the speed of light (in the relativistic limit when 
$m_{0}\gg 1$). They are of no interest for what is the intention to show.
The third term, which is analogous to the interference term, should be
analyzed in details. In the phase space its contribution is 
\begin{eqnarray*}
\rho _{int}(y,p,t) &=&e^{-\frac{p^{2}}{\Delta ^{2}}}\int dk\;\left[
1+w(p-k)w(p+k)\right] \left[ e^{-\frac{(k-m_{0})^{2}}{\Delta ^{2}}}+e^{-%
\frac{(k+m_{0})^{2}}{\Delta ^{2}}}\right] \\
&&\;e^{2iky-ie(p+k)t+ie(p-k)t}
\end{eqnarray*}

If $m_{0}$ is small (and so is $\Delta $ by assumption) then the function $%
w(k)$ is small and can be neglected. Also one can write $e(k)\approx 1+\frac{%
1}{2}k^{2}$, in which case the interference term in the phase space density
is 
\[
\rho _{int}(y,p,t)=e^{-\frac{p^{2}}{\Delta ^{2}}-\Delta ^{2}(x-pt)^{2}}\cos %
\left[ 2m_{0}(x-pt)\right] 
\]
Its typical form is shown in Figure 2 for the parameters $m_{0}=0.05$ and $%
\Delta =0.01$, and for two time instants: $t=0$ (left figure) and $t=5000$
(right figure). The main feature of the interference term is that its
modulus is independent of time and $m_{0}$. Furthermore it changes its shape
in unison with the rate at which the two main peaks in the phase space
density separate from each other.

\FRAME{ftbpFU}{4.7599in}{2.9196in}{0pt}{\Qcb{Interference term in the
initial phase space density (left figure) for the probability that is
localized around y=0 on the y-axes and if it has two isolated maxima along
the momentum axes. After the initial instant the probability on the y-axes
splits into two, and in the non-relativistic dynamics the interference term
in the phase space density adapts its shape instantaneously to the new
configuration (right figure). }}{\Qlb{fig2}}{fig2.wmf}{\special{language
"Scientific Word";type "GRAPHIC";display "USEDEF";valid_file "F";width
4.7599in;height 2.9196in;depth 0pt;original-width 6.1981in;original-height
3.8294in;cropleft "0";croptop "1";cropright "1";cropbottom "0";filename
'fig2.wmf';file-properties "XNPEU";}}

On the other hand if $m_{0}$ is large (but the width $\Delta $ is again
small), and because most of contribution to the interference term comes from 
$k=\pm m_{0}$, one can write $e(k)\approx |k|+\frac{1}{2|k|}$ and $%
1+w(p-k)w(p+k)\sim 2m_{0}^{-1}$. This means that the interference term
diminishes in the limit $m_{0}\rightarrow \infty $, and it is approximately 
\[
\rho _{int}(y,p,t)\sim \frac{1}{m_{0}}e^{-\frac{p^{2}}{\Delta ^{2}}-\Delta
^{2}x^{2}}\cos \left[ 2m_{0}\left( x-\frac{p}{m_{0}}t\right) \right] 
\]
which is valid in the time interval $t\ll m_{0}^{4}\Delta ^{-3}$. For longer
times it diminishes as $t^{-1}$. Besides being small contribution in the
overall phase space density the interference term has additional feature
that indicates its dependence on the time it takes the correlation to have
effect on it (the two receding peaks travel at nearly the speed of light).
The exponential term is ''frozen'' meaning that it is time independent, and
the oscillatory term is changing but with a great time lag, in fact it is
nearly constant also. If the two receding peaks are stopped by, say, a
potential step, then the situation would be similar to the two slit problem.
However, at that instant the interference part of the phase space density
would be very small in amplitude and not having the adequate shape. By
stopping the two peaks the phase space density would redistribute itself in
order to match this situation but it is obvious that process would take some
time. This indicates that the interference term in the two slits setup does
not form itself instantaneously but in reality it takes some time, and
therefore its source is in a physical process.

\section{Zero point energy}

Among the quantum effects is the so-called \textit{zero point energy}, which
is the lowest possible of all stationary states of a particle in a potential
(ground state energy). There are several reasons why it is in this group of
effects, but the one with the greatest weight comes from the interpretation
of the energy of the stationary states. According to the standard approach
to the quantum-classical relationship the correspondence between the quantum
and classical stationary states is only possible in the limit $\hbar
\rightarrow 0$, or equivalently for large quantum numbers (Bohr`s principle
of correspondence). As a consequence, it can be shown relatively easily from
the WKB approximation that the energy of stationary states have the
following interpretation. If position of a particle is random then the
modulus of its momentum is not, it is determined from the energy
conservation law. In other words, if $E_{0}$ is the energy of a stationary
state then the phase space density should parametrize, in the classical
limit, as 
\begin{equation}
\rho (\vec{r},\vec{p})\;=\;\delta \left( \frac{p^{2}}{2m}+V-E_{0}\right) =%
\frac{1}{\sqrt{2m(E_{0}-V)}}\delta (p-p^{\prime }\;)  \label{ro1}
\end{equation}
Indeed this limit is approached, on average, for the stationary states with
large quantum numbers. This means that for the ground state there is not
even approximate agreement between this classical limit and the quantum
ground state, as shown in Figure 3. Classical probability curve (dotted
line) has singularity at the points where momentum of the particle is zero.
Large portion of the quantum probability, however, is outside these
classical bounds, and the standard interpretation is that this is due to
tunneling, and therefore classical interpretation is not possible.

If the analysis starts from the Liouville equation (\ref{liouv}) then the
stationary solutions are obtained by requiring that $\partial _{t}\rho =0$,
which has a general solution in the form $\rho =F\left( \frac{p^{2}}{2m}%
+V\right) $, where $F$ is any function that has finite norm. Therefore there
is infinite number of stationary states, but one particular is 
\begin{equation}
\rho (x,p)=e^{-a\left( \frac{p^{2}}{2m}+\frac{1}{2}m\omega ^{2}x^{2}\right) }
\label{ro2}
\end{equation}
for, say, a one dimensional harmonic oscillator. The constant $a$ is
arbitrary and can be fixed by requiring that the average energy of the
oscillator is equal to the energy of the ground state of the quantum
oscillator, i.e. 
\[
<E>=\int dp\;\frac{p^{2}}{2m}\;Q(p)\;+\;\int dx\;\frac{1}{2}m\omega
^{2}x^{2}\;P(x)=E_{0} 
\]
The result for the probability density $P(x)$ is identical curve as in
Figure 3 for the quantum solution.

\FRAME{ftbpFU}{3.1315in}{3.4515in}{0pt}{\Qcb{Probability for the ground
state of the harmonic oscillator, if its energy is $E_{0}$. Quantum
calculation is shown by the curve $P_{qv}(x)$ and the WKB by $P_{WKB}(x)$.
The latter is treated in the standard interpretation as the classical limit
of quantum dynamics. The probability beyond the classical turning points is
interpreted as tunneling. }}{\Qlb{fig3}}{fig3.wmf}{\special{language
"Scientific Word";type "GRAPHIC";display "USEDEF";valid_file "F";width
3.1315in;height 3.4515in;depth 0pt;original-width 18.9749in;original-height
27.8418in;cropleft "0";croptop "1";cropright "1";cropbottom "0";filename
'fig3.WMF';file-properties "XNPEU";}}

The major departure from the standard classical analysis is in the
interpretation of what the energy of the stationary states represents. The
difference is summarized in the two expressions for the phase space
densities, (\ref{ro1}) and (\ref{ro2}). Thus according to the solution that
is based on the Liouville equation, both the coordinate and momentum of
particle are chosen randomly, according to the prescribed phase space
distribution. For each pair of these points in the phase space the
appropriate energy of the particle does not coincide with the energy of the
ground state, but on average it is equal to it. Therefore, the points in
Figure 3 that appear to be in the classical forbidden region, and hence
being interpreted as tunneling, are in fact manifestation of entirely
classical effect. The points that are considering tunneling express
probability of finding particle with the energy that is larger than $E_{0}$,
which according to (\ref{ro2}) is not zero. In the phase space this is
manifestly clear but from the solution of the Schroedinger equation (\ref
{schr}) it is not, because one works with the function that is the average
over all momentum part of the phase space. This in essence is the meaning of
the comment at the end of the previous section.

Previous analysis was entirely classical, and the identity with the quantum
result is accidental because the choice of the function $F$ was arbitrary,
however, the analysis explains the physical content of the zero point
energy. The only contribution of the uncertainty principle is to select from
the functions $F$ only those that satisfy it, and as it turns out there is
only one.

\section{Tunneling}

\label{sec:tun} The effect of tunneling is one of the most intriguing
quantum effects \cite{tunn}, but its proper understanding requires careful
analysis. As it has already been indicated what is considered to be the
tunneling effect in the case of the zero point energy it is in fact a
classical effect. It is manifestation of the phase space density when it is
averaged over the momentum subspace. However, that is not a typical
tunneling effect, more appropriately it is described in a scattering of
particle on a potential barrier. Discussion of this example starts by
considering a much simpler one, which is scattering of a particle from an
infinitely high potential barrier.

In this example the particle moves in the space on the negative x axes, and
the barrier is positioned at $x=0.$ If initial probability density $P_{0}(x)$
is given, and the initial probability current $j_{0}(x)$, then the problem
to find time evolution of the phase space density $\rho (x,p,t)$, and from
that the probability density $P(x,t)$, is well defined. There are several
reasons why the problem cannot be solved in the same way as for the
interference on two slits, which was described in the previous section. One
is that the potential is not of the parabolic type, which is a necessary
condition that the equation (\ref{schr}) is derived from the parametrization
(\ref{w}). Derivation of the equation (\ref{schr}) does not appear a
necessary condition for solving time evolution of the phase space density
because this is done by propagating trajectories in a potential. The
weakness of this argument is that by simple solution of this kind the
uncertainty principle may not be satisfied at arbitrary times, although at
the initial instant it is by the parametrization (\ref{w}). Therefore to
ensure that this is the case the function $f$ in the initial (\ref{w})
should incorporate all the information about the potential, including the
possible boundary conditions, and this is achieved by solving the equation (%
\ref{schr}). This argument appears non-physical because there is no reason
why at the initial instant the particle, if it is well localized away from
the barrier, should ''know'' of its existence. The same argument was applied
for the existence of the interference term in the two slits experiment, and
the answer was that this is an artifact of the non-relativistic theory,
where infinite velocities are possible. At the initial instant one can
ensure that the particle does not have ''contact'' with the barrier, but due
to the infinite dispersion of momenta at any short time after the particle
will be everywhere in the space, and therefore will know of the barrier. It
can be shown, but not elaborated here, that if the relativistic theory is
used, with the restriction that the probability density is zero outside
certain boundary, then indeed the particle does not know of the existence of
the barrier until this boundary reaches it. Because of that the boundary
propagates as in the traditional classical mechanics, i.e. no quantum
effects are observed. Therefore the initial (\ref{w}) should be determined
by using time independent solutions of the equation (\ref{schr}) as the
basis in which $P_{0}(x)$ is represented. However, there is a problem, as
mentioned earlier, the equation (\ref{schr}) cannot be derived for the
potentials other than harmonic. There is solution to this difficulty for
potentials of the step-like character, and the infinite barrier is of this
kind, which is to replace potential with the boundary condition. In other
words, one can neglect the potential barrier and treat the particle as being
free on the whole x axes. The barrier is mimicked by imposing the boundary
condition on the probability and the probability current at $x=0$ by
demanding that at each instant they should be equal to zero. The net effect
is the same, although the physics of the problem is not. The difference is
that physics demands that for $x>0$ the probability $P(x,t)$ is zero, while
from the imposed boundary condition this is not necessarily the case. In
other words, the problem is solved mathematically formally, but the physical
content should be extracted.

The initial function $f$, from which the initial phase space density is
calculated, is therefore obtained from a linear combination (again the units
are used in which $m=\hbar =1)$%
\[
f(x)=\frac{1}{2i}\int dk\;A(k)\;\left( e^{ikx}-e^{-ikx}\right) 
\]
where the amplitude $A(k)$ should be determined from $P_{0}(x)$ and $%
j_{0}(x) $. The plane waves $e^{\pm ikx}$ are solutions of the stationary
equation (\ref{schr}) and their combination ensures that the proper boundary
condition at $x=0$ is satisfied. For simplicity the initial conditions are 
\begin{equation}
P_{0}(x)=e^{-\frac{(x-x_{0})^{2}}{\Delta ^{2}}}\;\;\;;\;\;\;j_{0}(x)=p_{0}%
\;P_{0}(x)  \label{initp0}
\end{equation}
where $x_{0}$ is negative and chosen so that $|x_{0}|\gg \Delta $. It can be
shown that (constant pre-factors are omitted for convenience) 
\begin{equation}
A(k)=e^{-(k-p_{0})^{2}\Delta ^{2}/2-ikx_{0}}  \label{ampla}
\end{equation}
in which case the initial phase space density is 
\begin{eqnarray}
\rho _{0}(x,p) &=&\int_{-\infty }^{\infty }dq\;e^{2ipq}\;f^{*}(x+q)f(x-q)
\label{robarr} \\
&=&e^{-\frac{(x-x_{0})^{2}}{\Delta ^{2}}-(p-p_{0})^{2}\Delta ^{2}}+e^{-\frac{%
(x+x_{0})^{2}}{\Delta ^{2}}-(p+p_{0})^{2}\Delta ^{2}}-2\cos \left[
2(p_{0}x-px_{0})\right] e^{-\frac{x^{2}}{\Delta ^{2}}-p^{2}\Delta ^{2}} 
\nonumber
\end{eqnarray}
which consists of three terms: one centered around $x_{0}$ and $p_{0}$, the
other around $-x_{0}$ and $-p_{0}$, and the third around $x=p=0$ which
represents the interference term. The structure of the phase space density
is very similar to the one in the problem with two slits. However, the
interest here is not to discuss the effect of interference but to note one
important property of the momentum distribution, which has important
repercussions for understanding the tunneling effect.

At any later time the phase space density is given by (\ref{ro2t}), and the
probability for the coordinates is obtained from 
\[
P(x,t)=\int_{-\infty }^{\infty }dp\;\rho _{0}(x-pt,p) 
\]
which is not zero for $x>0$, but for $x<0$ it coincides with the true
solution of the problem. The physical solution is therefore obtained by
disregarding the probability for $x>0$. However, the momentum distribution
is obtained by integrating the physical solution for $f$, which is zero for $%
x>0$, but the phase space density (\ref{robarr}) was derived under the
assumption that this function extends over the whole x axes. Therefore, by
formally calculating the integral 
\[
Q(p,t)=\int_{-\infty }^{0}dx\;\rho _{0}(x-pt,p) 
\]
to obtain the momentum distribution is not a legitimate procedure. One needs
to extract $f(x,t)$ from the probability $P(x,t)$ and the probability
current $j(x,t)$, and then obtain $Q(p,t)$ from (\ref{ft}) and (\ref{pqq}).
This is the price one pays by replacing the potential barrier with the
boundary condition. For the case under discussion the function $f(x,t)$ is
relatively easily extracted, and the result is 
\[
f(x,t)=\frac{1}{\sqrt{\Delta ^{2}+it}}\left[ e^{\frac{-\Delta
^{2}(x-x_{0}-p_{0}t)^{2}+2i(\Delta ^{4}p_{0}-x_{0}t)x}{2(\Delta ^{4}+t^{2})}%
}-e^{\frac{-\Delta ^{2}(x+x_{0}+p_{0}t)^{2}-2i(\Delta ^{4}p_{0}-x_{0}t)x}{%
2(\Delta ^{4}+t^{2})}}\right] 
\]
The momentum function is 
\[
g(p,t)=\int_{-\infty }^{0}dx\;f(x,t)e^{-ipx} 
\]
which is not given explicitly because it is a rather lengthy expression,
instead a typical probability $Q(p,t)$ (solid line) is shown in Figure 4 for
four typical times: initial instant (a), just before the maximum of the
Gaussian probability reaches the barrier ($t_{0}=|x_{0}|/p_{0}$) (b), just
after this instant (c) and long after that (d). Its shape is not what one
would expect from the intuitive reasoning, and which is based on the
fundamental law of classical mechanics that any change in momentum is caused
by force.

\FRAME{ftbpFU}{2.9983in}{3.2534in}{0pt}{\Qcb{Various stages of scattering on
the infinite barrier as observed in the momentum space. At initial instant
(a) the momentum of the particle is centered around a positive value. Just
before (b) and after (c) the maximum of the probability on the coordinate
axes reaches the barrier the momentum distribution widens. The dotted line
is the momentum distribution from the traditional classical calculation.
Long after the collision (d) the momentum distribution is centered around a
negative value, and it is mirror image of the initial (a). }}{\Qlb{fig4}}{%
fig4.wmf}{\special{language "Scientific Word";type "GRAPHIC";display
"USEDEF";valid_file "F";width 2.9983in;height 3.2534in;depth
0pt;original-width 19.1314in;original-height 25.2153in;cropleft "0";croptop
"1";cropright "1";cropbottom "0";filename 'fig4.WMF';file-properties
"XNPEU";}}

In this particular example the force is of a special kind, it only changes
the sign of the momentum, and therefore the width of $Q(p,t)$ should not
change in time because the modulus of the momentum is constant. The expected
probability is the following: in the beginning particle moves towards the
barrier and its momentum distribution is centered around $p_{0}$, and after
long time the particle moves away from the barrier and its momentum
distribution is centered around $-p_{0}$. These two probabilities should be
the mirror images of each other. At any other time, in particular around $%
t=t_{0}$, the function $Q(p,t)$ is a combination of the two extreme cases.
This is indeed the case if $Q(p,t)$ is calculated from the phase space
distribution in which the interference term in (\ref{robarr}) is neglected
(from the traditional classical mechanics). The resulting probability is
shown in Figure 4 by the dotted line, which deviates considerably from the
probability when the uncertainty principle is implemented. The essence of
the difference is in the change of the modulus of the momentum, which in
traditional classical explanation it is attributed to the action of a force.
However, there is no such force only the infinite barrier. In fact the
momentum distribution changes dramatically between the initial instant and $%
t\sim t_{0}$, which is shown by calculating $g(p,t)$ for large $p$. In this
limit (the non-essential factors are omitted) 
\[
g(p,t)\sim \frac{1}{p^{2}}e^{\frac{-\Delta ^{2}(x_{0}+p_{0}t)^{2}}{2(\Delta
^{4}+t^{2})}} 
\]
which means that the modulus of the function changes from $g(p,t)=\exp
(-p^{2}\Delta ^{2}/2)$ to $g(p,t)\sim p^{-2}$ between these two instants.
The widening effect of the momentum distribution has no source in the
dynamics, because it is entirely the consequence of the change in the width
of the probability in the coordinate. In other words, as the width of the
probability in the coordinate of the particle changes the distribution of
the momenta also changes, meaning that no classical dynamics can explain
this effect because it is not caused by the action of a force.

The widening effect of the momentum distribution can be tested by assuming
that the barrier is not infinitely high, say it has the value $V_{0}$. One
consequence of the finite height is that all the phase space density for
which $p>\sqrt{2V_{0}}$ would ''leak'' into the half space $x>0$ and will
manifest itself as the non-zero probability $P(x,t)$. However, for a very
high barrier the estimate of this probability, under the assumption that
there is no effect due to the widening of the momentum distribution, gives 
\[
P(x,t)\sim e^{-2V_{0}\Delta ^{2}} 
\]
which is negligible small. Therefore, the prediction is, which is based on
considering only the initial distribution of momenta, that the probability
for particle to get over the barrier is negligible. On the other hand, if
widening is taken into account then the estimate of the probability $P(x,t)$
is obtained by first calculating the function $f(x,t)$ from 
\[
f(x,t)=\int dp\;e^{ipx}\;g(p,t)\sim e^{\frac{-\Delta ^{2}(x_{0}+p_{0}t)^{2}}{%
2(\Delta ^{4}+t^{2})}}\int dk\;\;\frac{e^{ikx}}{k^{2}+2V_{0}} 
\]
where $p$ was replaced by $\sqrt{k^{2}+2V_{0}}$ so that it is explicitly
taken into account that $p>2V_{0}$. By evaluating the integral one gets the
estimate 
\begin{equation}
P(x,t)\sim e^{\frac{-\Delta ^{2}(x_{0}+p_{0}t)^{2}}{(\Delta ^{4}+t^{2})}%
}\;e^{-2x\sqrt{2V_{0}}}  \label{cltunn}
\end{equation}
which has two important features. One, the probability for over the barrier
transmission is incomparable larger than the estimate based only on the
initial distribution of momenta, however, this happens when $t\sim
|x_{0}/p_{0}|$. Second, the probability decays exponentially for increasing $%
x$, but there is no time dependence of it, except in the factor that
indicates arrival of the incident probability. In other words, the
probability for over the barrier transmission spreads instantaneously in the
whole $x>0$ half space. This, again, as an artifact of the non-relativistic
dynamics, which can be shown by the relativistic treatment of this problem
(because of its rather lengthy treatment the details are not given here).

The effect for over the barrier transmission, and the exponential dependence
of its probability with the coordinate, can be tested by making the
potential zero at some distance $x=\delta >0$. If there is relatively large
probability to observe the particle in the space $x>\delta $ then that would
be direct test of the uncertainty principle, because this is the only way to
explain its appearance. It would be called a paradox, and it is called the 
\textit{tunneling effect, }because the only conclusion from the initial
conditions is that such events are not possible (or with the negligible
probability), and classical dynamics cannot account for such a large
dispersion of momenta. However, it will be shown that once uncertainty
principle is implemented the tunneling effect has classical explanation as
the over the barrier transmission.

If the effect of tunneling is manifestation of the uncertainty principle,
and not dynamics of the particle, then the question is whether there is any
meaning in saying that the solutions of (time dependent) Schroedinger
equation can be obtained by solving classical equations of motion. The
answer is affirmative because inability to describe the change in the
momentum distribution, which is due to the uncertainty principle, is
replaced by the unusual initial phase space distribution. The meaning of
this will be demonstrated on the more exact calculation for the over the
barrier transmission, i.e. the tunneling probability on a step potential.
Scattering on the step potential has been analyzed in details, and it was
showed that classical and quantum calculations for the probability $P(x,t)$
produce identical results. Therefore the analysis of this example will not
be analyzed in details, only the essentials points in the part that is
relevant for this discussion.

The idea is, as discussed previously, to replace the potential by the
boundary condition, but for the step potential its implementation is
somewhat more elaborate than in the case of the infinite barrier.
Complication is caused by the fact that the phase space density is not
confined to the space $x<0$, but it is also transmitted into the space $x>0$%
. Because of that it is necessary to analyze two separate sets of
trajectories: one in the zero potential and the other in the potential $%
V_{0} $. This means that if the same idea as for the infinite barrier is
used then two separate phase space densities should be defined.: one when
the potential is zero ($\rho _{1}$) and the other ($\rho _{2}$) when it has
the value $V_{0}$. Both are defined on the whole x axes, but the phase space
density $\rho _{1}(x,p,t)$ is only meaningful in the space $x<0$ while $\rho
_{2}(x,p,t)$ in the space $x>0$. The initial conditions are set on the
negative x axes, which means that the phase space density $\rho _{1}(x,p,t)$
is defined from them. On the other hand, the phase space density $\rho
_{2}(x,p,t)$ does not have direct relationship to them, only indirectly
through the boundary condition at $x=0$ for the two quantities: the
probability densities $P_{1}(x,t)$ and $P_{2}(x,t)$, and the probability
currents $j_{1}(x,t)$ and $j_{2}(x,t)$ that are defined for the phase space
densities $\rho _{1}(x,p,t)$ and $\rho _{2}(x,p,t)$, respectively. At the
boundary it is required that $P_{1}(x,t)$ $=P_{2}(x,t)$ and $j_{1}(x,t)=$ $%
j_{2}(x,t)$.

The details of deriving the phase space densities are omitted, because that
was shown elsewhere \cite{bos8}. The analysis is relatively complicated for
the review, and so only the final result will be cited, for the particular
case of interest: the tunneling. The initial phase space density $\rho
_{2}^{0}(x,p)$ is given by 
\[
\rho _{2}^{0}(x,p)=\;16\func{Re}\left\{ \int dk\,A(k)A^{*}(k_{p}^{-})\;\frac{%
k\left[ 2p-K(k)\right] }{\left[ k+K(k)\right] \left[ k+K(k_{p}^{-})\right] }%
\;e^{2ix[K(k)-p]}\right\} \, 
\]
where $K(k)=\sqrt{k^{2}-2V_{0}}$. The variable $k_{p}^{-}$ is defined as 
\[
k_{p}^{-}\;=\pm \;\left[ K(k)-2p+i\sqrt{2V_{0}}\right] ^{1/2}\left[ K(k)-2p-i%
\sqrt{2V_{0}}\right] ^{1/2} 
\]
where the sign is selected from the requirement that $\func{Im}[K(k_{p}^{-})%
]<0$. The integration path is in the upper half of the complex k-plane,
which has important feature to avoid two Riemann cuts that are defined
there, and preferably it should go through the stationary point of the phase
of the integrand (this is particularly important for the study of
tunneling). For the specific case when $\sqrt{2V_{0}}\gg p_{0}$ the initial
phase space density is approximately (the non-essential factors are again
omitted) 
\[
\rho _{2}^{0}(x,p)\sim e^{-2x\sqrt{2V_{0}}}\;\func{Re}\left[ e^{-2ixp}\int
dk\,A(k)A^{*}(k_{p}^{-})\;k\;\right] 
\]
which appears not to make physical sense: it is unbounded on the negative x
axes. This is very unpleasant because at later time the phase space density
is 
\[
\rho _{2}(x,p,t)=\rho _{2}^{0}(x-pt,p)\sim e^{-2(x-pt)\sqrt{2V_{0}}}\;\func{%
Re}\left[ e^{-2i(x-pt)p}\int dk\,A(k)A^{*}(k_{p}^{-})\;k\right] 
\]
and in the space $x>0$, where by definition it is meaningful, its amplitude
increases without bounds in the limit $t\rightarrow \infty $. Therefore it
could be rejected as non-physical. However, it should be recalled that it is
the probability $P_{2}(x,t)$ that has physical meaning, and this should be
finite for all time. In other words, from the physics of the problem it
should follow that 
\begin{equation}
\stackunder{t\rightarrow \infty }{\lim }P_{2}(x,t)=\stackunder{t\rightarrow
\infty }{\lim }\int dp\;\rho _{2}^{0}(x-pt,p)=0  \label{limp2}
\end{equation}
For the amplitude (\ref{ampla}) this can be explicitly proved by evaluating
two integrals, in the variables $k$ and $p$, by the stationary phase method,
where the phase of the integrand is 
\[
\vartheta (p,k)=-\frac{1}{2}(k_{p}^{-}-p_{0})^{2}\Delta ^{2}+ik_{p}^{-}x_{0}-%
\frac{1}{2}(k-p_{0})^{2}\Delta ^{2}-ikx_{0}-2(x-pt)\left( \sqrt{2V_{0}}%
+ip\right) 
\]
The set 
\[
\partial _{p}\vartheta (p,k)=0\;\;\;\;\;;\;\;\;\;\;\partial _{k}\vartheta
(p,k)=0 
\]
defines the stationary points, which are obtained by first making the
replacement $p=r/(\Delta ^{2}\sqrt{2V_{0}})$ and then in the equations
retain the leading terms in the powers of $V_{0}$. The solution of this
approximate set of equations is (the details are not shown, because
obtaining it is straightforward but relatively lengthy) 
\[
k_{st}=\frac{p_{0}\Delta ^{2}-ix_{0}}{\Delta ^{2}+it}\;\;\;\;;\;\;\;p_{st}=%
\Delta ^{2}\left( -p_{0}\Delta ^{4}+tx_{0}\right) \frac{x_{0}+tp_{0}}{\sqrt{%
2V_{0}}\left( \Delta ^{4}+t^{2}\right) ^{2}} 
\]
The probability $P_{2}(x,t)$ is therefore 
\[
P_{2}(x,t)\sim \;\func{Re}\left[ e^{\vartheta (p_{st},k_{st})}\right] 
\]
which, it can be shown, is equal to (\ref{cltunn}), and also to the solution
from the quantum treatment. The limit $t\rightarrow \infty $ is finite, but
not zero, however it is very small. The fact that this limit is not zero is
an artifact of the choice for the initial probability, which will be
discussed shortly. Therefore, despite the fact that the phase space density
increases without bounds in the space $x<0$ the probability $P_{2}(x,t)$ has
all prerogatives to be physically acceptable.

The stationary value $\;p_{st}$ for the momentum plays the role of the
average momentum $p_{0}$ for the free particle, but in this case it is
measured with respect to the potential $V_{0}$, and its initial value is $%
p_{st}=-\frac{p_{0}x_{0}}{\sqrt{V_{0}}}$, i.e it is very small. One confirms
this by calculating the probability current $j_{2}(x,t)$ in the space $x>0$,
and from the definition (\ref{curr}) the velocity of the particle is $%
v_{tunn}=j_{2}(x,t)/P_{2}(x,t)$. This calculation produces the identity $%
v_{tunn}=p_{st}$. Therefore, if one can talk about the \textit{tunneling
velocity} of the particle then this would be $v_{tunn}$, and it follows that
it is very small.

As it was shown classical mechanics describes tunneling but not because it
is result of dynamics, but because of the specific initial phase space
density that results from implementing the uncertainty principle. From this
phase space density, and by using solutions of classical equations of
motion, one describes tunneling effect, and the result is the same as by
solving time dependent Schroedinger equation. One aspect of this solution is
quite intriguing, and needs to be understood properly. Time dependence of
the tunneling probability (\ref{cltunn}) has a very specific form: time
variation of the probability at $x=0$ is instantaneously transmitted to all
points $x>0$. This means that at some point $x=\delta $ the time variation
of the probability is identical with that at the point $x=0$. If the
potential is cut at $x=\delta $ then propagation of the probability in the
space $x>\delta $ is the same as for a free particle. In fact it has the
same time dependence as if one chooses for it the initial $P_{0}(x)$, and as
if there is no gap. The only difference is that the amplitude of this
probability is scaled by the factor $e^{-2\delta \sqrt{2V_{0}}}$. The net
effect is that the probability travels from the space $x<0$ to $x>\delta $
as if there is no gap, i.e. as if the tunneling velocity is infinite. This
finding directly contradicts what had been shown before, that the tunneling
velocity is very small. The controversy can be resolved by noting one
important physical aspect of tunneling: the interval within which the
tunneling probability is significant is of the order $x_{tunn}\sim
(2V_{0})^{-1/2}$. This means that the time it takes for the particle to
travel this distance, at the velocity $v_{tunn}$, is 
\[
t_{tunn}=\frac{x_{tunn}}{v_{tunn}}\sim \frac{\left( \Delta ^{4}+t^{2}\right)
^{2}}{\Delta ^{2}\left( -p_{0}\Delta ^{4}+tx_{0}\right) \left(
x_{0}+tp_{0}\right) }\sim -\frac{\Delta ^{2}}{p_{0}x_{0}} 
\]
which is independent of $V_{0}$. In the last step time dependence was
neglected. Physical circumstances require that $|x_{0}|\gg \Delta $, and
also that during the time of scattering the shape of the probability is
nearly constant, which means that $t\ll \Delta ^{2}$. From the
characteristic collision time $t\sim |x_{0}|/p_{0}$ it follows that $%
p_{0}=Mx_{0}/\Delta ^{2}$, where $M$ is a large number. The tunneling time
is then 
\[
t_{tunn}\sim \left( \frac{\Delta }{x_{0}}\right) ^{2}\frac{\Delta ^{2}}{M} 
\]
which is very short compared with the characteristic time variation of the
probability for the free particle. Indeed, if the tunneling time is
multiplied by $p_{0}$ then the distance that the free probability travels
during this interval is 
\[
x=p_{0}t_{tunn}\sim x_{0}\left( \frac{\Delta }{x_{0}}\right) ^{2}=\Delta 
\frac{\Delta }{x_{0}} 
\]
which is small. Therefore, the reason why the probability has time
dependence (\ref{cltunn}) is that the tunneling velocity, although in
absolute magnitude is very small, is sufficiently large so that any change
at one end is transmitted ''instantaneously'' to the other.

There are other questions in connection with the tunneling effect, e.g. how
the specific form of the spatial dependence (\ref{cltunn}) is formed, but
they cannot be answered without considering more precise theoretical model.
In particular to answer these questions the initial probability (\ref{initp0}%
) is not adequate, because it would be more accurate to work with the one
that is strictly zero outside certain boundary. In connection with this one
should also use relativistic theory, because any cutoff in the probability
on the x axes makes distribution of momenta very wide, which also includes
relativistic values, i.e. for a given momentum $p$ the velocity of particle $%
p/\sqrt{1+p^{2}}$ is nearly the speed of light. However, considering these
issues would require more extensive discussion, but results would not
contribute in an essential way to the understanding of the tunneling effect.

\section{Conclusion}

The aim of the previous discussion was to show that the two basic quantum
effects: interference and tunneling, can be explained and quantitatively
described by formulating dynamics of a particle from the classical
principles, with addition of the uncertainty principle. The results are
identical to those obtained from quantum mechanics, and by that it is meant
solutions of Schroedinger equation. The two approaches are different ways of
seeing the same effects, in many respects analogous to analyzing the motion
of classical particles from either the Newton`s equations of motion or from
the Lagrange principle of least action. Or, solving the harmonic oscillator
problem starting from matrix mechanics or from the differential equation.
The advantage of analyzing quantum effects from the classical principles is
that one works in the phase space, and therefore sees the problem with
additional degrees of freedom. In this respect more information is available
about the system, which is lost if one only works in, say, the coordinate
space. The last is characteristic of quantum mechanics, and although one can
switch between the coordinate and momentum spaces, one never works in both
at the same time. Crudely speaking, quantum mechanics works with the
averaged quantities in one of the phase space coordinates and because of
that one easily makes erroneous conclusions. This is best observed in the
analysis of the zero point energy in one of the previous sections. From
quantum mechanics one makes conclusion that relatively large portion of the
probability for the zero point energy is due to tunneling, but in fact its
shape is explained entirely classically.

There was attempt to overcome this drawback of quantum mechanics by
formulating \textit{quantum phase space density}, which would enable to
study dynamics of the particle in all the phase space variables. The
transform that extends quantum dynamics into the phase space is the Wigner
function $w(\vec{r},\vec{p},t)$, which was mentioned in Section 2. For the
wave function $\psi (\vec{r},t)$ it is defined through the property 
\[
|\psi (\vec{r},t)|^{2}\;=\;\int d^{3}p\,\,w(\vec{r},\vec{p}%
,t)\;\;\;\;;\;\;\;\;|\phi (\vec{p},t)|^{2}\;=\;\int d^{3}r\,\;w(\vec{r},\vec{%
p},t) 
\]
where $\phi (\vec{p},t)$ is the wave function in the momentum space. This,
however, is the only connection with the phase space density that was used
throughout the paper. Namely, the Wigner function, or the quantum phase
space density, does not satisfy any simple equation, and definitely not the
Liouville equation, except in the special case of the harmonic-type forces.
The Wigner function is used as the mean to study the limit $\hbar
\rightarrow 0$ of quantum mechanics, and as the proof that this is classical
mechanics it is shown that Liouville equation is obtained. Based on such
arguments the quantum phase space density was used in the study of the
quantum-classical relationship \cite
{muga1,muga2,balazs,heller,lee,bonci,berry,bonasera,
royer,ghosh,smerzi,kasperkovitz}. The quantum phase space density is
mentioned in the context of this work because it is the base of a dilemma:
is classical mechanics the limit of quantum when $\hbar \rightarrow 0$ or is
quantum mechanics derived from classical by taking into account the
uncertainty principle? This point was discussed elsewhere \cite{bos8} and
therefore will not be discussed here in details. The resime of this
discussion is that by strictly taking the limit $\hbar \rightarrow 0$ in
quantum mechanics one obtains classical but with a special property, in
which the phase space density is parametrized as (\ref{ro1}). On the other
hand, by starting from classical mechanics one obtains correct result for,
say, the ground state of harmonic oscillator.

The disadvantage of working in the phase space (or from classical
principles) is that solving problems is more difficult, and often not
straightforward. As discussed on the example of tunneling the essential new
feature that the uncertainty principle introduces into dynamics is that the
change in the momentum is not necessarily caused by the action of a force
(it can be called \textit{non-dynamic effect}). Therefore, action of a force
on the particle is not sufficient to reproduce quantum results, one needs to
take into account that momentum changes from the other cause, the
implementation of the uncertainty principle. The exception is harmonic
force, for which it can be shown not to affect the phase space density in a
way that would change the momentum distribution that cannot be explained by
the force itself. This can be demonstrated by solving scattering problem on
inverted parabolic potential $V(x)=-\frac{1}{2}wx^{2}$. Given initial
conditions $x_{i}$ and p$_{i}$ for classical trajectory in this potential
its time dependence is 
\[
x=x_{i}\,cosh(\omega t)+\frac{p_{i}}{\omega }\,sinh(\omega
t)\;\;\;;\;\;p=p_{i}\,cosh(\omega t)+\omega x_{i}\,sinh(\omega t)\; 
\]
and the phase space density $\rho (x,p,t)$, if it is initial $\rho _{0}(x,p)$%
, is 
\[
\rho (x,p,t)=\rho _{0}\left[ x\;cosh(\omega t)-\frac{p}{\omega }%
\,sinh(\omega t),-p\;\,cosh(\omega t)+\omega x\,sinh(\omega t)\right] \; 
\]
From this phase space density one calculates the probability $P(x,t)$ from
the definition (\ref{pq}). This is classical solution for arbitrary $\rho
_{0}(x,p)$, and specifically if the initial conditions (\ref{initp0}) are
chosen then the result is 
\[
P(x,t)=\frac{1}{\sqrt{\pi \Delta _{t}^{2}}}\,e^{-(x-x_{t})^{2}/\Delta
_{t}^{2}} 
\]
where 
\begin{eqnarray*}
\Delta _{t} &=&\Delta \left[ cosh^{2}(\omega t)\,+\,\frac{sinh^{2}(\omega t)%
}{\omega ^{2}\Delta ^{4}}\right] ^{1/2} \\
x_{t} &=&x_{0}\,cosh(\omega t)\,+\,\frac{p_{0}}{\omega }\,sinh(\omega t)
\end{eqnarray*}
The same result is obtained if the probability $P(x,t)$ is calculated from
quantum mechanics, i.e. by solving Schroedinger equation for the time
evolution of the wave function. This result is independent of the particular
choice of the initial conditions, but for the Gaussian type (\ref{initp0})
it has convenient analytic form.

For other than harmonic potentials the contribution of the non-dynamic
effects may be significant, but sometime negligible. However, the problem
can be avoided by working with the step like potentials, as mentioned in
Section 2, in which case the change in potential is replaced by the boundary
conditions on the phase space density. By doing that one also includes the
non-dynamic effects into account. The procedure is exact, but the result for
the time evolution of the phase space density is relatively complicated. The
solution simplifies considerably if one is only interested in the time
evolution of the probability $P(x,t)$, when it is sufficient to solve the
equation (\ref{schr}), the procedure that is valid for any potential. That
it is a correct one can be demonstrated on one example. It will be assumed
that the potential is a delta function at the origin, i.e. $V(x)=W_{0}\delta
(x)$, and the initial conditions are (\ref{initp0}). The potential divides
the space into $x<0$ and $x>0$, and in each one the particle moves in zero
potential. Time evolution of the phase space density $\rho ^{>}(x,p,t)$ in $%
x>0$ can be thought to originate from some initial $\rho _{0}^{>}(x,p)$.
That initial phase space density is obtained by requiring that at $x=0$ the
phase space density $\rho ^{<}(x,p,t)$ for the space $x<0$ changes smoothly
into $\rho ^{>}(x,p,t)$. The initial phase space density for $\rho
^{<}(x,p,t)$ is obtained from (\ref{initp0}) by the same procedure as
described in the section on tunneling, with a slight modification due to the
fact that the phase space density penetrates into the space $x>0$. The
proper connection between the two phase space densities is ensured by the
proper choice of the function $f$ that enters the phase space density (\ref
{w}). It can be shown, without giving the details, that this function for
the two spaces is 
\[
f^{<}(x)=\int dk\;A(k)\left[ e^{ikx}+R\left( k\right) e^{-ikx}\right]
\;\;\;;\;\;\;f^{>}(x)=\int dk\;A(k)T(k)e^{ikx} 
\]
where the coefficients are 
\[
R(k)=-\frac{iW_{0}}{k+iW_{0}}\;\;\;;\;\;\;T(k)=\frac{k}{k+iW_{0}} 
\]
The phase space density $\rho _{0}^{>}(x,p)$ has analytic form if $%
W_{0}>>p_{0}$, in which case 
\[
\rho _{0}^{>}(x,p)\sim \frac{1}{W_{0}^{2}}e^{-\Delta ^{2}(p-p_{0})^{2}-\frac{%
1}{\Delta 2}(x-x_{0})^{2}}\left[ 2(x-x_{0})^{2}+2\Delta ^{4}p^{2}-\Delta ^{2}%
\right] 
\]
and its time evolution is $\rho ^{>}(x,p,t)=\rho _{0}^{>}(x-pt,p)$. The
probability $P(x,t)$ of finding the particle in the space $x>0$ (the
tunneling probability) is then (normalization is omitted) 
\[
P(x,t)=\int dp\;\rho ^{>}(x,p,t)=\frac{(x-x_{0})^{2}+\Delta ^{4}p_{0}^{2}}{%
W_{0}^{2}\left( \Delta ^{4}+t^{2}\right) ^{3/2}}\;e^{\frac{\Delta ^{2}}{%
\Delta ^{4}+t^{2}}(x-x_{0}-p_{0}t)^{2}} 
\]
which is exactly the same result as if the problem was solved by quantum
mechanics.

In conclusion one can say that the alternative formulation of dynamics of
particle, which is based on the classical principles with the amendment of
the uncertainty principle, gives identical results as the original
formulation in terms of the wave-particle dualism. In other words,
Schroedinger equation is derived from these classical principles, which was
confirm in the analysis of two quantum effects: interference and tunneling.

\end{document}